# *Three-dimensional multiple-relaxation-time lattice Boltzmann model for convection heat transfer in porous media at the REV scale*

Q. Liu, Y.-L. He

*Key Laboratory of Thermo-Fluid Science and Engineering of Ministry of Education, School of Energy and Power Engineering, Xi'an Jiaotong University, Xi'an, Shaanxi, 710049, China*

**Abstract**

In this paper, a three-dimensional (3D) multiple-relaxation-time (MRT) lattice Boltzmann (LB) model is presented for convection heat transfer in porous media at the representative elementary volume (REV) scale. The model is developed in the framework of the double-distribution-function (DDF) approach: an MRT-LB model of the density distribution function with the D3Q19 lattice (or D3Q15 lattice) is proposed to simulate the flow field based on the generalized non-Darcy model, while an MRT-LB model of the temperature distribution function with the D3Q7 lattice is proposed to simulate the temperature filed. The present model is employed to simulate mixed convection flow in a porous channel and natural convection in a cubical porous cavity. The numerical results demonstrate the effectiveness and accuracy of the present model in solving 3D convection heat transfer problems in porous media. The numerical results also demonstrate that the present model is approximately second-order accuracy in space. In addition, an enthalpy-based DDF-MRT model for 3D solid-liquid phase change with convection heat transfer in porous media is also presented.

**Keywords**: Lattice Boltzmann method; Multiple-relaxation-time (MRT); Porous media; Convection heat transfer; Three-dimensional (3D); REV scale.

## 1. Introduction

Fluid flow and heat transfer in porous media have attracted considerable attention due to their fundamental nature and broad range of applications in many fields of science and engineering [1-4]. Over the last several decades, various numerical methods have been developed to study fluid flow and heat transfer in porous media. The lattice Boltzmann (LB) method [5-14], as a mesoscopic numerical method originated from the lattice-gas automata (LGA) method [15], has achieved great success in simulating fluid flow and heat transfer in porous media due to its kinetic background [16-35].

The LB models for fluid flow and heat transfer in porous media can be generally classified into two categories: the pore scale method [16-21] and the REV scale method [22-35]. Shortly after its emergence in the late 1980s, the LB method was applied to study 3D incompressible flows in a random medium by Succi et al [16]. In the pore scale method [16-21], fluid flow in the pores of the medium is directly modeled by the standard LB method. By using the no-slip bounce-back rule, the interaction between the fluid and the solid matrix can be handled efficiently. The main advantage of the pore scale method is that the detailed local information (e.g., permeability) of the flow in the pores can be obtained, which can be used to investigate macroscopic relations (e.g., the Darcy's law). Moreover, some fundamental issues such as medium variability and scale dependency can be assessed quantitatively [17].

In the REV scale method [22-35], an additional term accounting for the presence of a porous medium is added into the LB equation based on some semi-empirical models (e.g., Darcy model, Brinkman-extended Darcy model, and generalized non-Darcy model). In this method, the statistical properties (e.g., porosity, permeability, inertia coefficient) of the medium are incorporated into the model directectly with the detailed structure being ignored, and thus the detailed local information of the flow in the pores are usually unavailable. However, by using appropriate models of the porous medium, reasonable results can be produced by this method. What's more, the REV scale LB method can be used to study fluid flow and heat transfer in porous media systems of large size. Based on the generalized non-Darcy model (also called the Brinkman-Forchheimer-extended Darcy model) [36], Guo and Zhao [25] proposed a generalized LB model for studying incompressible flows in porous media. In the generalized LB model, the influence of the porous matrix is considered by including the porosity into the equilibrium distribution function, and adding a forcing term to the LB equation to account for the linear (Darcy's term) and nonlinear (Forchheimer's term) drag forces of the porous matrix. Subsequently, the generalized LB model was extended to simulate convection heat transfer in porous media [27, 28]. After nearly two decades of development, the REV scale LB method has been developed into an accurate and efficient numerical tool for studying fluid flow and heat transfer problems in porous media systems. In the literature [32-35], some efforts have also been made to develop thermal LB models for studying solid-liquid phase change with convection heat transfer in porous media. By combing the enthalpy method with the thermal LB method, the moving solid-liquid interface can be traced by updating the enthalpy without imposing hydrodynamic or thermal boundary conditions.

It is noted that most of the existing REV scale thermal LB models for heat transfer problems [27-35] in porous media without/with solid-liquid phase change are limited to two-dimensional (2D) cases. Little attention has been paid to study heat transfer problems in 3D porous media systems via 3D REV scale thermal LB method. As a promising numerical tool for engineering applications, it is desirable to develop 3D REV scale thermal LB model to generalize our understanding of fluid flow and heat transfer processes in 2D porous media systems to those in 3D cases. Hence, the aim of this paper is to develop a 3D REV scale thermal LB model using the MRT approach considering that the MRT collision operator [37, 38] is superior over its BGK counterpart [7]. The model is developed in the framework of the DDF approach: a D3Q19-MRT model (or D3Q15-MRT model) is proposed to simulate the flow field based on the generalized non-Darcy model, while a D3Q7-MRT model is proposed to simulate the temperature filed. Numerical simulations of mixed convection flow in a porous channel and natural convection in a cubical porous cavity are carried out to validate the 3D DDF-MRT model.

The rest of this paper is organized as follows. The macroscopic governing equations are briefly described in Section 2. In Section 3, the 3D DDF-MRT model for convection heat transfer in porous media is presented in detail. Numerical results and some discussions are given in Section 4. Finally, a brief conclusion is made in Section 5.

## 2. Macroscopic governing equations

For fluid flow and convection heat transfer in an isotropic and rigid porous medium, based on the generalized non-Darcy model, the macroscopic governing equations under local thermal equilibrium (LTE) condition can be written as follows [4, 36]:

$$\nabla \cdot \mathbf{u} = 0 \tag{1}$$

$$\frac{\partial \mathbf{u}}{\partial t} + (\mathbf{u} \cdot \nabla)\left(\frac{\mathbf{u}}{\phi}\right) = -\frac{1}{\rho_0}\nabla(\phi p) + v_e \nabla^2 \mathbf{u} + \mathbf{F} \tag{2}$$

$$\sigma\frac{\partial T}{\partial t} + \mathbf{u} \cdot \nabla T = \nabla \cdot (\alpha_e \nabla T) \tag{3}$$

where $\mathbf{u}$, $T$, and $p$ are the volume-averaged fluid velocity, temperature, and pressure, respectively, $\rho_0$ is the mean fluid density, $\phi$ is the porosity, $v_e$ is the effective kinematic viscosity, $\sigma$ is the heat capacity ratio $\sigma = [\phi\rho_f c_{pf} + (1-\phi)\rho_m c_{pm}]/(\rho_f c_{pf})$ (ratio between mean heat capacity of the mixture and fluid heat capacity), and $\alpha_e$ is the effective thermal diffusivity $\alpha_e = \phi\alpha_f + (1-\phi)\alpha_m$

($\alpha_f$ and $\alpha_m$ are thermal diffusivities of the fluid and porous matrix, respectively). $c_p$ is the specific heat, and the subscripts $f$ and $m$ refer to the properties of fluid and solid matrix, respectively.

$\mathbf{F} = (F_x, F_y, F_z)$ denotes the total body force induced by porous matrix and external force fields, which can be expressed as [27, 39]

$$\mathbf{F} = -\frac{\phi v}{K}\mathbf{u} - \frac{\phi F_\phi}{\sqrt{K}}|\mathbf{u}|\mathbf{u} + \phi \mathbf{G} \qquad (4)$$

where $K$ is the permeability of the porous medium, $v$ is the kinematic viscosity of the fluid ($v$ is not necessarily the same as $v_e$), and $|\mathbf{u}| = \sqrt{u_x^2 + u_y^2 + u_z^2}$, in which $u_x$, $u_y$, and $u_z$ are components of the fluid velocity $\mathbf{u}$ in the $x$-, $y$-, and $z$-directions, respectively. The first and second terms on the right hand side of Eq. (4) are linear (Darcy's term) and nonlinear (Forchheimer's term) drag forces of the porous matrix. Without the nonlinear drag force term ($F_\phi = 0$), Eq. (2) reduces to the Brinkman-extended Darcy equation. According to Boussinesq approximation, the body force $\mathbf{G}$ is given by

$$\mathbf{G} = -\mathbf{g}\beta(T - T_0) + \mathbf{a} \qquad (5)$$

where $\mathbf{g}$ is the gravitational acceleration, $\beta$ is the thermal expansion coefficient, $T_0$ is the reference temperature, and $\mathbf{a}$ is the acceleration induced by other external force fields.

Based on Ergun's experimental investigations [40], the inertia coefficient $F_\phi$ and the permeability $K$ can be expressed as [41]

$$F_\phi = \frac{1.75}{\sqrt{150\phi^3}}, \quad K = \frac{\phi^3 d_p^2}{150(1-\phi)^2} \qquad (6)$$

where $d_p$ is the solid particle diameter.

Convection heat transfer in porous media governed by Eqs. (1)-(3) is characterized by several dimensionless parameters: the Darcy number $Da$, the Rayleigh number $Ra$, the Reynolds number $Re$ (for mixed convection flow), the Prandtl number $Pr$, the viscosity ratio $J$, and the thermal

diffusivity ratio $\Gamma$, which are defined as follows

$$Da = \frac{K}{L^2}, \quad Ra = \frac{g\beta\Delta T L^3}{\nu\alpha}, \quad Re = \frac{Lu_c}{\nu}, \quad Pr = \frac{\nu}{\alpha}, \quad J = \frac{\nu_e}{\nu}, \quad \Gamma = \frac{\alpha_e}{\alpha} \qquad (7)$$

where $L$ is the characteristic length, $u_c$ is the characteristic velocity, and $\alpha$ is the thermal diffusivity of the fluid.

## 3. 3D DDF-MRT model for convection heat transfer in porous media

The MRT method was proposed by d'Humières [37] in 1992, which is an important extension of the relaxation LB method developed by Higuera et al. [6]. In the LB community, it has been widely accepted that the MRT collision operator [37, 38] is superior over its BGK counterpart [7]. In recent years, several DDF-MRT models have been proposed to simulate heat transfer problems without porous media in two [42-44] and three [45, 46] dimensions. In our previous studies [30, 34], the DDF-MRT method has also been proposed to simulate heat transfer problems in porous media. In this section, a 3D DDF-MRT model for convection heat transfer in porous media is presented, which can be viewed as an extension to our previous studies.

### 3.1 MRT-LB model for the flow field

For the flow field, the MRT-LB equation with an explicit treatment of the forcing term can be written as [30, 47, 48]

$$f_i(\mathbf{x}+\mathbf{e}_i\delta_t, t+\delta_t) = f_i(\mathbf{x},t) - \tilde{\Lambda}_{ij}\left(f_j - f_j^{eq}\right)\Big|_{(\mathbf{x},t)} + \delta_t\left(S_i - 0.5\tilde{\Lambda}_{ij}S_j\right)\Big|_{(\mathbf{x},t)} \qquad (8)$$

where $f_i(\mathbf{x},t)$ is the density distribution function, $f_i^{eq}(\mathbf{x},t)$ is the equilibrium density distribution function, $\tilde{\mathbf{\Lambda}} = \mathbf{M}^{-1}\mathbf{\Lambda}\mathbf{M}$ is the collision matrix (**M** is the transformation matrix, **Λ** is the relaxation matrix), and $S_i$ is the forcing term.

Through the transformation matrix **M**, the collision process of the MRT-LB equation (8) can be executed in the moment space $\mathbb{M}$:

$$\mathbf{m}^{*}(\mathbf{x},t) = \mathbf{m}(\mathbf{x},t) - \mathbf{\Lambda}(\mathbf{m}-\mathbf{m}^{eq})\big|_{(\mathbf{x},t)} + \delta_{t}\left(\mathbf{I}-\frac{\mathbf{\Lambda}}{2}\right)\tilde{\mathbf{S}} \qquad (9)$$

The streaming process is still executed in the velocity space $\mathbb{V}$:

$$f_{i}(\mathbf{x}+\mathbf{e}_{i}\delta_{t}, t+\delta_{t}) = f_{i}^{*}(\mathbf{x},t) \qquad (10)$$

where $\mathbf{f}^{*} = \mathbf{M}^{-1}\mathbf{m}^{*}$. The bold-face symbols $\mathbf{m}$, $\mathbf{m}^{eq}$ and $\tilde{\mathbf{S}}$ denote b-dimensional column vectors:

$$\mathbf{m} = |m\rangle = \mathbf{M}\mathbf{f}, \quad \mathbf{m}^{eq} = |m^{eq}\rangle = \mathbf{M}\mathbf{f}^{eq}, \quad \tilde{\mathbf{S}} = |\tilde{S}\rangle = \mathbf{M}\mathbf{S} \qquad (11)$$

in which $\mathbf{f} = |f\rangle$, $\mathbf{f}^{eq} = |f^{eq}\rangle$, and $\mathbf{S} = |S\rangle$. For brevity, the Dirac notation $|\cdot\rangle$ is adopted to denote a b-dimensional column vector, e.g., $|m\rangle = (m_{0}, m_{1}, \ldots, m_{b-1})^{\mathrm{T}}$. For the flow filed, the D3Q15 and D3Q19 lattices can be used in the MRT-LB model, which leads to the D3Q15-MRT and D3Q19-MRT models. The transformation matrices [49] are given in Appendix A. In what follows, the two MRT models for the flow field are introduced.

### 3.1.1 D3Q15-MRT model

The 15 discrete velocities $\{\mathbf{e}_{i}|i=0, 1, \ldots, 14\}$ of the D3Q15 lattice are given by (see Fig. 1)

$$\mathbf{e} = c\begin{bmatrix} 0 & 1 & -1 & 0 & 0 & 0 & 0 & 1 & -1 & 1 & -1 & 1 & -1 & 1 & -1 \\ 0 & 0 & 0 & 1 & -1 & 0 & 0 & 1 & 1 & -1 & -1 & 1 & 1 & -1 & -1 \\ 0 & 0 & 0 & 0 & 0 & 1 & -1 & 1 & 1 & 1 & 1 & -1 & -1 & -1 & -1 \end{bmatrix} \qquad (12)$$

where $c = \delta_{x}/\delta_{t}$ is the lattice speed with $\delta_{t}$ and $\delta_{x}$ being the discrete time step and lattice spacing in x-direction, respectively. The lattice speed $c$ is set to be 1 ($\delta_{x} = \delta_{t}$) in this work.

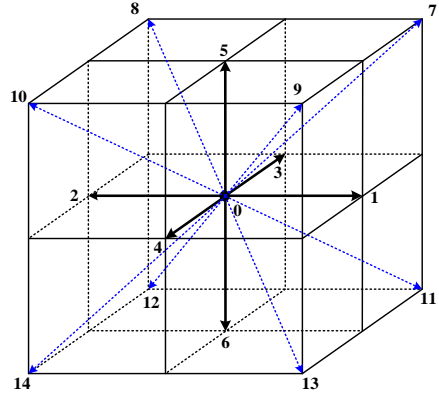

Fig. 1. Discrete velocities of the D3Q15 lattice.

The moment vector $\mathbf{m}$ is defined as

$$\mathbf{m} = \left(\rho, e, \varepsilon, \tilde{j}_x, q_x, \tilde{j}_y, q_y, \tilde{j}_z, q_z, 3p_{xx}, p_{ww}, p_{xy}, p_{yz}, p_{zx}, m_{xyz}\right)^{\mathrm{T}} \tag{13}$$

where

$$\tilde{j}_x = j_x - \tfrac{\delta_t}{2}\rho F_x, \quad \tilde{j}_y = j_y - \tfrac{\delta_t}{2}\rho F_y, \quad \tilde{j}_z = j_z - \tfrac{\delta_t}{2}\rho F_z \tag{14}$$

with $j_x = \rho u_x$, $j_y = \rho u_y$, and $j_z = \rho u_z$. For incompressible flows considered in this work, the incompressible approximation is adopted, i.e., $\rho = \rho_0 + \delta\rho \approx \rho_0$ ($\delta\rho$ is the density fluctuation), then the flow momentum is approximated by $\mathbf{J} = (j_x, j_y, j_z) \approx \rho_0 \mathbf{u}$.

The equilibrium moments $m_i^{eq}$ for the non-conserved moments $m_i$ ($i \neq 0,3,5,7$) are given by

$$e^{eq} = -\rho + \frac{\rho_0 |\mathbf{u}|^2}{\phi}, \quad \varepsilon^{eq} = \beta_1 \rho + \beta_2 \frac{\rho_0 |\mathbf{u}|^2}{\phi}, \quad q_x^{eq} = -\frac{7}{3}\rho_0 u_x, \quad q_y^{eq} = -\frac{7}{3}\rho_0 u_y$$

$$q_z^{eq} = -\frac{7}{3}\rho_0 u_z, \quad 3p_{xx}^{eq} = \rho_0 \frac{2u_x^2 - u_y^2 - u_z^2}{\phi}, \quad p_{ww}^{eq} = \rho_0 \frac{u_y^2 - u_z^2}{\phi}$$

$$p_{xy}^{eq} = \rho_0 \frac{u_x u_y}{\phi}, \quad p_{yz}^{eq} = \rho_0 \frac{u_y u_z}{\phi}, \quad p_{zx}^{eq} = \rho_0 \frac{u_z u_x}{\phi}, \quad m_{xyz}^{eq} = 0 \tag{15}$$

where $\beta_1$ and $\beta_2$ are free parameters.

The diagonal relaxation matrix $\Lambda$ is given by

$$\Lambda = \mathrm{diag}\left(s_\rho, s_e, s_\varepsilon, s_j, s_q, s_j, s_q, s_j, s_q, s_\nu, s_\nu, s_\nu, s_\nu, s_\nu, s_m\right) \tag{16}$$

The components of the forcing term $\tilde{\mathbf{S}}$ are given as follows:

$$\tilde{S}_0 = 0, \quad \tilde{S}_1 = 2\rho_0 \frac{\mathbf{u} \cdot \mathbf{F}}{\phi}, \quad \tilde{S}_2 = -10\rho_0 \frac{\mathbf{u} \cdot \mathbf{F}}{\phi}, \quad \tilde{S}_3 = \rho_0 F_x, \quad \tilde{S}_4 = -\frac{7}{3}\rho_0 F_x, \quad \tilde{S}_5 = \rho_0 F_y, \quad \tilde{S}_6 = -\frac{7}{3}\rho_0 F_y,$$

$$\tilde{S}_7 = \rho_0 F_z, \quad \tilde{S}_8 = -\frac{7}{3}\rho_0 F_z, \quad \tilde{S}_9 = 2\rho_0 \frac{(2u_x F_x - u_y F_y - u_z F_z)}{\phi}, \quad \tilde{S}_{10} = 2\rho_0 \frac{(u_y F_y - u_z F_z)}{\phi}$$

$$\tilde{S}_{11} = \rho_0 \frac{u_x F_y + u_y F_x}{\phi}, \quad \tilde{S}_{12} = \rho_0 \frac{u_y F_z + u_z F_y}{\phi}, \quad \tilde{S}_{13} = \rho_0 \frac{u_x F_z + u_z F_x}{\phi}, \quad \tilde{S}_{14} = 0 \tag{17}$$

The equilibrium distribution function $f_i^{eq}$ in the velocity space is given by ($\beta_1 = 1$, $\beta_2 = -5$)

$$f_i^{eq} = w_i \left\{ \rho + \rho_0 \left[ \frac{\mathbf{e}_i \cdot \mathbf{u}}{c_s^2} + \frac{(\mathbf{e}_i \cdot \mathbf{u})^2}{2\phi c_s^4} - \frac{|\mathbf{u}|^2}{2\phi c_s^2} \right] \right\} \tag{18}$$

where $w_0 = 2/9$, $w_{1\sim 6} = 1/9$, $w_{7\sim 14} = 1/72$, and $c_s = 1/\sqrt{3}$ is the lattice sound speed.

### 3.1.2 D3Q19-MRT model

The 19 discrete velocities $\{\mathbf{e}_i | i = 0, 1, \ldots, 18\}$ of the D3Q19 lattice are given by (see Fig. 2)

$$\mathbf{e} = c \begin{bmatrix} 0 & 1 & -1 & 0 & 0 & 0 & 0 & 1 & -1 & 1 & -1 & 1 & -1 & 1 & -1 & 0 & 0 & 0 & 0 \\ 0 & 0 & 0 & 1 & -1 & 0 & 0 & 1 & 1 & -1 & -1 & 0 & 0 & 0 & 0 & 1 & -1 & 1 & -1 \\ 0 & 0 & 0 & 0 & 0 & 1 & -1 & 0 & 0 & 0 & 0 & 1 & 1 & -1 & -1 & 1 & 1 & -1 & -1 \end{bmatrix} \tag{19}$$

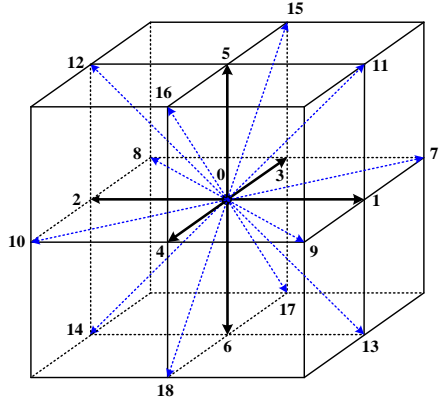

Fig. 2. Discrete velocities of the D3Q19 lattice.

The moment vector $\mathbf{m}$ is defined as

$$\mathbf{m} = \left(\rho, e, \varepsilon, \tilde{j}_x, q_x, \tilde{j}_y, q_y, \tilde{j}_z, q_z, 3p_{xx}, 3\pi_{xx}, p_{ww}, \pi_{ww}, p_{xy}, p_{yz}, p_{zx}, m_x, m_y, m_z\right)^{\mathrm{T}} \tag{20}$$

The equilibrium moments $m_i^{eq}$ for the non-conserved moments $m_i$ ($i \neq 0, 3, 5, 7$) are given by

$$e^{eq} = -11\rho + 19 \frac{\rho_0 |\mathbf{u}|^2}{\phi}, \quad \varepsilon^{eq} = \beta_1 \rho + \beta_2 \frac{\rho_0 |\mathbf{u}|^2}{\phi}, \quad q_x^{eq} = -\frac{2}{3}\rho_0 u_x$$

$$q_y^{eq} = -\frac{2}{3}\rho_0 u_y, \quad q_z^{eq} = -\frac{2}{3}\rho_0 u_z, \quad 3p_{xx}^{eq} = \rho_0 \frac{2u_x^2 - u_y^2 - u_z^2}{\phi}$$

$$3\pi_{xx}^{eq} = 3\beta_3 p_{xx}^{eq}, \quad p_{ww}^{eq} = \rho_0 \frac{u_y^2 - u_z^2}{\phi}, \quad \pi_{ww}^{eq} = \beta_3 p_{ww}^{eq}, \quad p_{xy}^{eq} = \rho_0 \frac{u_x u_y}{\phi}$$

$$p_{yz}^{eq} = \rho_0 \frac{u_y u_z}{\phi}, \quad p_{zx}^{eq} = \rho_0 \frac{u_z u_x}{\phi}, \quad m_x^{eq} = m_y^{eq} = m_z^{eq} = 0 \tag{21}$$

where $\beta_1$, $\beta_2$, and $\beta_3$ are free parameters.

The diagonal relaxation matrix $\mathbf{\Lambda}$ is given by

$$\mathbf{\Lambda} = \mathrm{diag}\left(s_\rho, s_e, s_\varepsilon, s_j, s_q, s_j, s_q, s_j, s_q, s_v, s_\pi, s_v, s_\pi, s_v, s_v, s_v, s_m, s_m, s_m\right) \tag{22}$$

The components of the forcing term $\tilde{\mathbf{S}}$ are given as follows:

$$\tilde{S}_0 = 0, \quad \tilde{S}_1 = 38\rho_0 \frac{\mathbf{u} \cdot \mathbf{F}}{\phi}, \quad \tilde{S}_2 = -11\rho_0 \frac{\mathbf{u} \cdot \mathbf{F}}{\phi}, \quad \tilde{S}_3 = \rho_0 F_x, \quad \tilde{S}_4 = -\frac{2}{3}\rho_0 F_x$$

$$\tilde{S}_5 = \rho_0 F_y, \quad \tilde{S}_6 = -\frac{2}{3}\rho_0 F_y, \quad \tilde{S}_7 = \rho_0 F_z, \quad \tilde{S}_8 = -\frac{2}{3}\rho_0 F_z$$

$$\tilde{S}_9 = 2\rho_0 \frac{(2u_x F_x - u_y F_y - u_z F_z)}{\phi}, \quad \tilde{S}_{10} = -\rho_0 \frac{(2u_x F_x - u_y F_y - u_z F_z)}{\phi}$$

$$\tilde{S}_{11} = 2\rho_0 \frac{u_y F_y - u_z F_z}{\phi}, \quad \tilde{S}_{12} = -\rho_0 \frac{u_y F_y - u_z F_z}{\phi}, \quad \tilde{S}_{13} = \rho_0 \frac{u_x F_y + u_y F_x}{\phi}$$

$$\tilde{S}_{14} = \rho_0 \frac{u_y F_z + u_z F_y}{\phi}, \quad \tilde{S}_{15} = \rho_0 \frac{u_x F_z + u_z F_x}{\phi}, \quad \tilde{S}_{16} = \tilde{S}_{17} = \tilde{S}_{18} = 0 \tag{23}$$

The equilibrium distribution function $f_i^{eq}$ in the velocity space is given by ($\beta_1 = 3$, $\beta_2 = -5.5$, $\beta_3 = -0.5$)

$$f_i^{eq} = w_i \left\{ \rho + \rho_0 \left[ \frac{\mathbf{e}_i \cdot \mathbf{u}}{c_s^2} + \frac{(\mathbf{e}_i \cdot \mathbf{u})^2}{2\phi c_s^4} - \frac{|\mathbf{u}|^2}{2\phi c_s^2} \right] \right\} \tag{24}$$

where $w_0 = 1/3$, $w_{1\sim 6} = 1/18$, $w_{7\sim 18} = 1/36$, and $c_s = 1/\sqrt{3}$ is the lattice sound speed.

The macroscopic fluid density $\rho$ and velocity $\mathbf{u}$ of the MRT-LB model are defined by

$$\rho = \sum_i f_i \tag{25}$$

$$\rho_0 \mathbf{u} = \sum_i \mathbf{e}_i f_i + \frac{\delta_t}{2} \rho_0 \mathbf{F} \tag{26}$$

The macroscopic fluid pressure $p$ is defined as $p = \rho c_s^2 / \phi$. Note that Eq. (26) is a nonlinear equation for the velocity $\mathbf{u}$. By introducing a temporal velocity $\mathbf{v}$, the macroscopic fluid velocity $\mathbf{u}$ can be calculated explicitly by

$$\mathbf{u} = \frac{\mathbf{v}}{l_0 + \sqrt{l_0^2 + l_1 |\mathbf{v}|}} \tag{27}$$

where

$$\mathbf{v} = \sum_{i=0}^{8} \mathbf{e}_i f_i / \rho_0 + \frac{\delta_t}{2} \phi \mathbf{G}, \quad l_0 = \frac{1}{2}\left(1 + \phi \frac{\delta_t}{2} \frac{v}{K}\right), \quad l_1 = \phi \frac{\delta_t}{2} \frac{F_\phi}{\sqrt{K}} \tag{28}$$

Through the Chapman-Enskog analysis [47] of the MRT-LB equation (8), the following macroscopic equations can be obtained:

$$\frac{\partial \rho}{\partial t} + \nabla \cdot (\rho \mathbf{u}) = 0 \tag{29}$$

$$\frac{\partial(\rho\mathbf{u})}{\partial t}+\nabla\cdot\left(\frac{\rho\mathbf{u}\mathbf{u}}{\phi}\right)=-\nabla(\phi p)+\nabla\cdot\mathbf{\Pi}+\rho\mathbf{F} \tag{30}$$

where

$$\mathbf{\Pi}=\rho v_e\left[\nabla\mathbf{u}+(\nabla\mathbf{u})^\mathsf{T}\right]+\rho\left(v_B-\frac{2}{3}v_e\right)(\nabla\cdot\mathbf{u})\mathbf{I} \tag{31}$$

is the shear stress tensor. In the incompressible limit, the macroscopic equations (29) and (30) reduce to the generalized Navier-Stokes equations (1) and (2). The effective kinematic viscosity $v_e$ and the bulk viscosity $v_B$ are defined as

$$v_e=c_s^2\left(\frac{1}{s_v}-\frac{1}{2}\right)\delta_t,\quad v_B=\frac{5-9c_s^2}{9}\left(\frac{1}{s_e}-\frac{1}{2}\right)\delta_t \tag{32}$$

**3.2 MRT-LB model for the temperature field**

For the temperature field, a new MRT-LB model is proposed in this subsection. the MRT-LB equation is given by

$$g_i(\mathbf{x}+\mathbf{e}_i\delta_t,t+\delta_t)-g_i(\mathbf{x},t)=-\left(\mathbf{N}^{-1}\mathbf{Q}\mathbf{N}\right)_{ij}\left(g_j-g_j^{eq}\right)\Big|_{(\mathbf{x},t)} \tag{33}$$

where $g_i(\mathbf{x},t)$ is the temperature distribution function, $g_i^{eq}(\mathbf{x},t)$ is the equilibrium temperature distribution function, $\mathbf{N}$ is the transformation matrix, and $\mathbf{Q}$ is the relaxation matrix.

Through the transformation matrix $\mathbf{N}$, the collision process of the MRT-LB equation (33) can be executed in the moment space $\mathbb{M}$:

$$\mathbf{n}^*(\mathbf{x},t)=\mathbf{n}(\mathbf{x},t)-\mathbf{Q}\left(\mathbf{n}-\mathbf{n}^{eq}\right)\Big|_{(\mathbf{x},t)} \tag{34}$$

where $\mathbf{n}=|n\rangle=\mathbf{N}\mathbf{g}$, $\mathbf{n}^{eq}=|n^{eq}\rangle=\mathbf{N}\mathbf{g}^{eq}$. The streaming process is executed in the velocity space $\mathbb{V}$:

$$g_i(\mathbf{x}+\mathbf{e}_i\delta_t,t+\delta_t)=g_i^*(\mathbf{x},t) \tag{35}$$

where $\mathbf{g}^*=\mathbf{N}^{-1}\mathbf{n}^*$.

As the temperature is regarded as a passive scalar, the D3Q7 lattice can be used in the MRT-LB model for the temperature field. The 7 discrete velocities $\{\mathbf{e}_i|i=0,1,\ldots,6\}$ of the D3Q7 lattice are given in Eq. (12) (see Fig. 1). In the present study, the transformation matrix $\mathbf{N}$ of the D3Q7-MRT

model is given by

$$\mathbf{N} = \begin{bmatrix} 1 & 1 & 1 & 1 & 1 & 1 & 1 \\ 0 & 1 & -1 & 0 & 0 & 0 & 0 \\ 0 & 0 & 0 & 1 & -1 & 0 & 0 \\ 0 & 0 & 0 & 0 & 0 & 1 & -1 \\ 0 & 1 & 1 & 1 & 1 & 1 & 1 \\ 0 & 1 & 1 & -1 & -1 & 0 & 0 \\ 0 & 1 & 1 & 0 & 0 & -1 & -1 \end{bmatrix} \quad (36)$$

The transformation matrix $\mathbf{N}$ is constructed based on the following basis vectors: $|1\rangle$, $|e_x\rangle$, $|e_y\rangle$, $|e_z\rangle$, $|e_x^2+e_y^2+e_z^2\rangle$, $|e_x^2-e_y^2\rangle$, $|e_x^2-e_z^2\rangle$. The equilibrium moments $\{n_i^{eq}\}$ are defined as

$$\mathbf{n}^{eq} = T(\sigma, u_x, u_y, u_z, \varpi\sigma, 0, 0)^{\mathrm{T}} \quad (37)$$

where $\varpi \in (0,1)$.

The relaxation matrix $\mathbf{Q}$ is given by

$$\mathbf{Q} = \mathrm{diag}(\zeta_T, \zeta_\alpha, \zeta_\alpha, \zeta_\alpha, \zeta_e, \zeta_\nu, \zeta_\nu) \quad (38)$$

$n_0$ is the only conserved moment and the temperature $T$ is computed by

$$\sigma T = n_0 = \sum_i g_i \quad (39)$$

For convection heat transfer without porous media, the equilibrium moments $\{n_i^{eq}\}$ are defined by $\mathbf{n}^{eq} = T(1, u_x, u_y, u_z, \varpi, 0, 0)^{\mathrm{T}}$, and the temperature $T$ is computed by $T = n_0 = \sum_i g_i$.

Through the Chapman-Enskog analysis of the MRT-LB equation (33), the following macroscopic equation can be obtained

$$\frac{\partial(\sigma T)}{\partial t} + \nabla\cdot(\mathbf{u}T) = \nabla\cdot\left\{\delta_t\left(\zeta_\alpha^{-1} - 0.5\right)\left[c_{sT}^2 \nabla(\sigma T) + \epsilon\partial_{t_1}(\mathbf{u}T)\right]\right\} \quad (40)$$

where $c_{sT} = \sqrt{\varpi/3}$ is the lattice sound speed of the D3Q7 model. In most cases, the deviation term $\nabla\cdot\left[\delta_t\left(\zeta_\alpha^{-1} - 0.5\right)\epsilon\partial_{t_1}(\mathbf{u}T)\right]$ in Eq. (40) can be neglected for incompressible thermal flows, then the temperature governing equation (3) can be recovered under the assumption that $\sigma$ does not change with time and varies slowly in space. The effective thermal diffusivity is $\alpha_e$ defined as

$$\alpha_e = \sigma c_{sT}^2 \left(\frac{1}{\zeta_\alpha} - \frac{1}{2}\right)\delta_t \quad (41)$$

The equilibrium temperature distribution function $g_i^{eq}$ ($\mathbf{g}^{eq} = \mathbf{N}^{-1}\mathbf{n}^{eq}$) in the velocity space is given by

$$g_i^{eq} = \omega_i T \left( \sigma + \frac{\mathbf{e}_i \cdot \mathbf{u}}{c_{sT}^2} \right) \quad (42)$$

where $\omega_0 = 1 - \varpi$, $\omega_{1\sim 6} = \varpi/6$. The inverse of $\mathbf{N}$ is given by

$$\mathbf{N}^{-1} = \begin{bmatrix} 1 & 0 & 0 & 0 & -1 & 0 & 0 \\ 0 & \frac{1}{2} & 0 & 0 & \frac{1}{6} & \frac{1}{6} & \frac{1}{6} \\ 0 & -\frac{1}{2} & 0 & 0 & \frac{1}{6} & \frac{1}{6} & \frac{1}{6} \\ 0 & 0 & \frac{1}{2} & 0 & \frac{1}{6} & -\frac{1}{3} & \frac{1}{6} \\ 0 & 0 & -\frac{1}{2} & 0 & \frac{1}{6} & -\frac{1}{3} & \frac{1}{6} \\ 0 & 0 & 0 & \frac{1}{2} & \frac{1}{6} & \frac{1}{6} & -\frac{1}{3} \\ 0 & 0 & 0 & -\frac{1}{2} & \frac{1}{6} & \frac{1}{6} & -\frac{1}{3} \end{bmatrix} \quad (43)$$

The D3Q7-MRT model can be extended to simulate solid-liquid phase change with convection heat transfer in porous media. In Appendix B, an enthalpy-based D3Q7-MRT model for solid-liquid phase change with convection heat transfer in porous media is briefly introduced.

## 4. Numerical results and discussions

In this section, numerical simulations of mixed convection flow in a porous channel, natural convection in a cubical porous cavity, and two-region conduction melting in a semi-infinite space are carried out to validate the present model.Unless otherwise specified, we set $\rho_0 = 1$, $\delta_t = 1$, $\delta_x = \delta_y = \delta_z = 1$, $c = 1$, $J = 1$, $\sigma = 1$, $\Gamma = 1$, and $\varpi = 1/2$ ($c_{sT}^2 = 1/6$). The free relaxation rates are chosen as follows: $s_\rho = s_j = 1$, $s_e = s_\varepsilon = s_q = 1.1$, $s_m = 1.2$ (D3Q15-MRT model); $s_\rho = s_j = 1$, $s_e = s_\varepsilon = s_q = s_\pi = 1.1$, $s_m = 1.2$ (D3Q19-MRT model); $\zeta_T = 1$, $\zeta_e = \zeta_v = 1.2$ (D3Q7-MRT model). The non-equilibrium extrapolation scheme [50] is employed to treat the velocity and temperature boundary conditions. For natural convection in a cubical porous cavity, the D3Q19-MRT model is employed to simulated the flow field.

### 4.1 Mixed convection flow in a porous channel

In this subsection, we first test the present model by simulating the mixed convection flow in a porous channel (see Fig. 3). The distance between the two parallel plates is $H$, the upper plate is hot ($T = T_h$) and moves along the x-direction with a uniform velocity $u_0$, while the bottom plate is cold ($T = T_c$). A constant normal flow of fluid is injected (with a uniform velocity $u_1$) through the bottom plate and is withdrawn at the same rate from the upper plate. Without the nonlinear drag force ($F_\phi = 0$), the flow at steady state is governed by the following equations [27, 30, 31]:

$$\frac{u_y}{\phi}\frac{\partial u_x}{\partial y} = v_e \frac{\partial^2 u_x}{\partial y^2} - \frac{\phi v}{K} u_x \tag{44}$$

$$\frac{1}{\rho_0}\frac{\partial p}{\partial y} = g\beta(T-T_0) - \frac{v}{K}u_y + a_y \tag{45}$$

$$u_y \frac{\partial T}{\partial y} = \nabla \cdot (\alpha_e \nabla T) \tag{46}$$

where $T_0 = (T_h + T_c)/2$ is the reference temperature, and $a_y$ is the external force in the y-direction:

$$a_y = \frac{v}{K}u_1 - g\beta\Delta T \left[\frac{\exp(yu_1/\alpha_e)-1}{\exp(Hu_1/\alpha_e)-1}\right] \tag{47}$$

The analytical solutions of Eqs. (44)-(46) are given by

$$u_x = u_0 \exp\left[\vartheta_1\left(\frac{y}{H}-1\right)\right]\frac{\sinh(\vartheta_2 \cdot y/H)}{\sinh(\vartheta_2)}, \quad u_y = u_1, \quad u_z = 0 \tag{48}$$

$$T = T_c + \Delta T \frac{\exp(Pr_e Re \cdot y/H)-1}{\exp(Pr_e Re)-1} \tag{49}$$

where $Re = Hu_1/v$ is the Reynolds number, $Pr_e = v/\alpha_e$ is the effective Prandtl number, $\Delta T = T_h - T_c$ is the temperature difference. The two parameters $\vartheta_1$ and $\vartheta_2$ in Eq. (48) are given by

$$\vartheta_1 = \frac{Re}{2\phi J}, \quad \vartheta_2 = \frac{1}{2\phi J}\sqrt{Re^2 + \frac{4\phi^3 J}{Da}} \tag{50}$$

In simulations, we set $Ra = 100$, $Pr = 1$ ($Pr_e = Pr$), $\phi = 0.6$, and $u_0 = u_1 = 0.01$. Periodic boundary conditions are imposed in the x- and z-directions, and a grid size of $N_x \times N_y \times N_z = 6 \times 32 \times 6$ is adopted. The relaxation rates $s_v$ and $\zeta_\alpha$ are determined by

$$s_v = \frac{1}{2} + \frac{JHu_1}{c_s^2 Re\delta_t}, \quad \zeta_\alpha^{-1} = \frac{1}{2} + \frac{\Gamma c_s^2 \left(s_v^{-1} - 0.5\right)}{J\sigma c_{sT}^2 Pr} \tag{51}$$

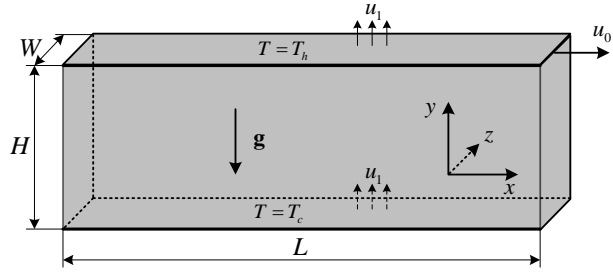

Fig. 3. Mixed convection flow in a porous channel.

In Figs. 4 and 5, the normalized velocity and temperature profiles for different Reynolds numbers and Darcy numbers are plotted and compared with the analytical solutions. As can be observed, the present results are in excellent agreement with the analytical solutions.

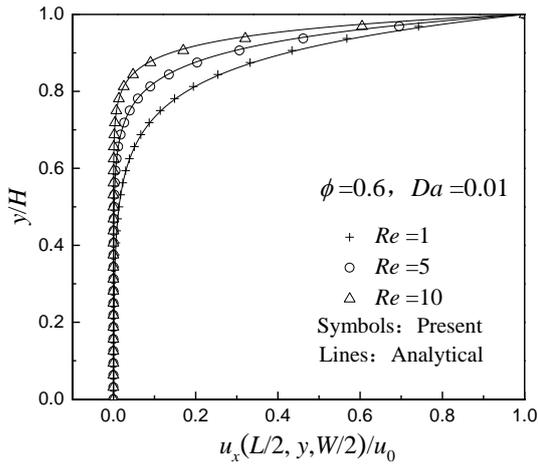

(a) velocity profiles

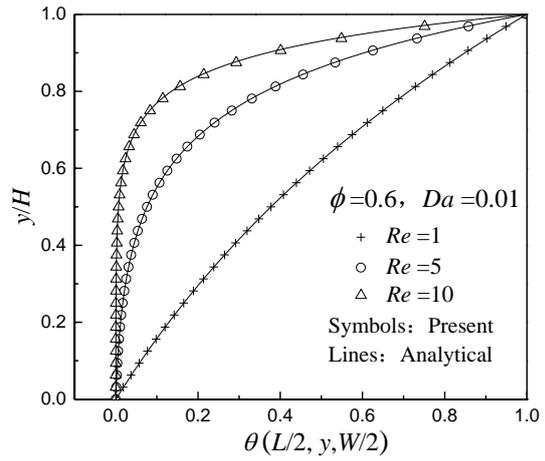

(b) temperature profiles

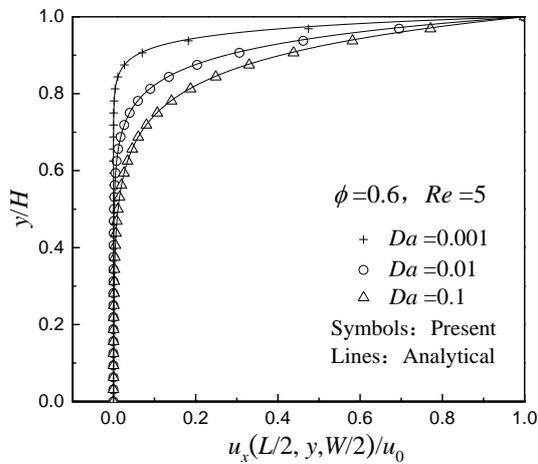

(c) velocity profiles

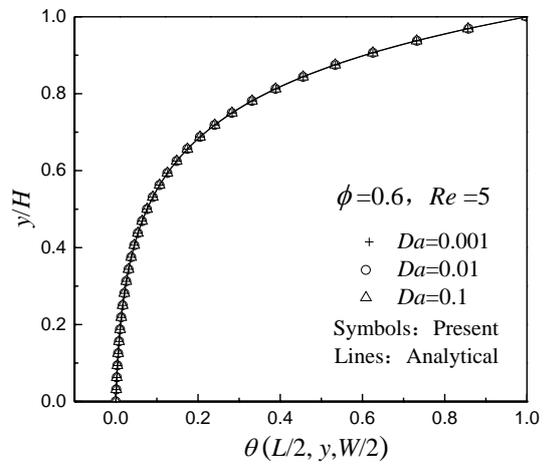

(d) temperature profiles

Fig. 4. Velocity and temperature profiles for different $Re$ and $Da$ (D3Q15-MRT model).

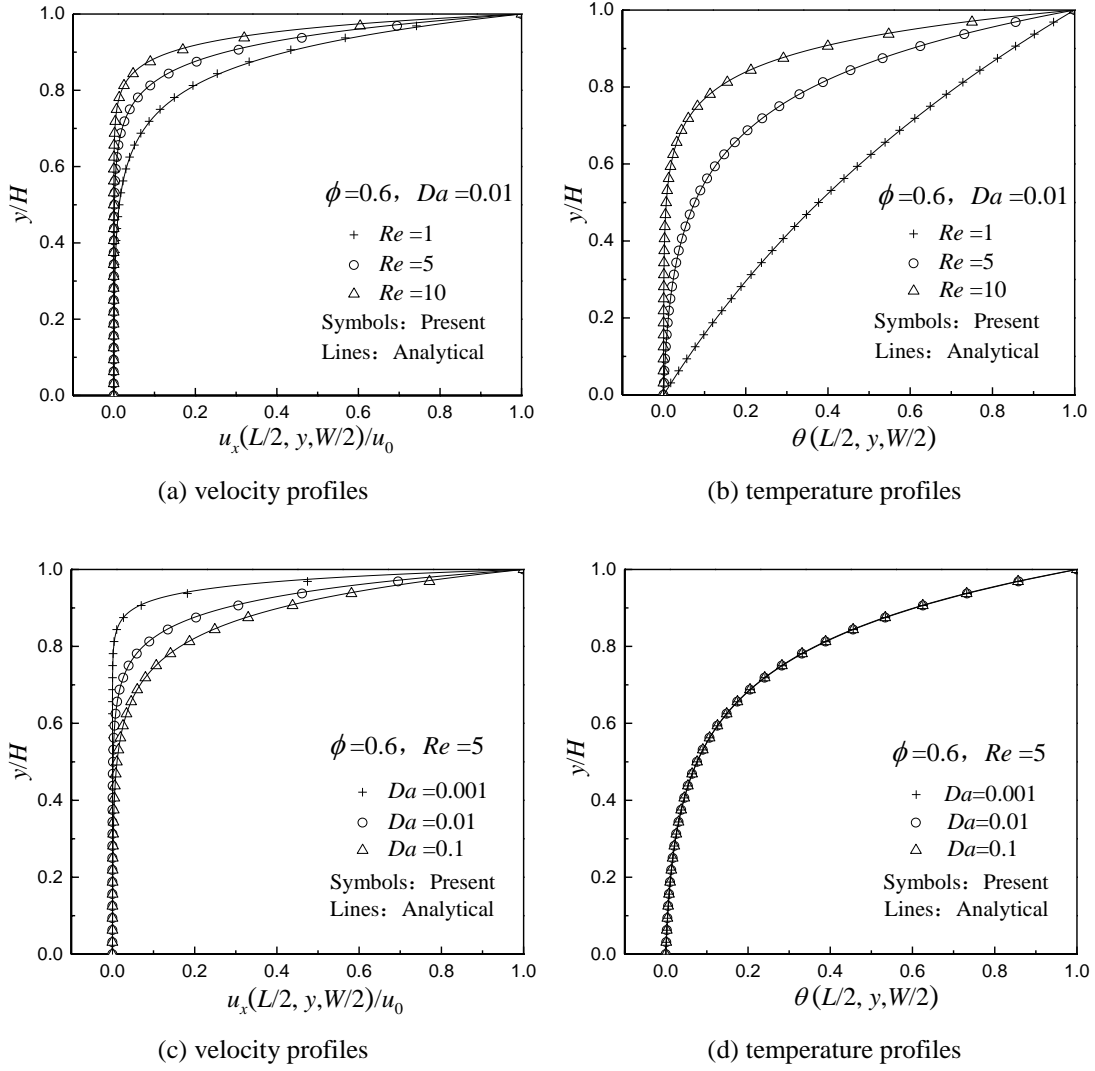

(a) velocity profiles

(b) temperature profiles

(c) velocity profiles

(d) temperature profiles

Fig. 5. Velocity and temperature profiles for different $Re$ and $Da$ (D3Q19-MRT model).

Numerical simulations are also carried out to evaluate the spatial accuracy of the present model. In simulations, we set $Ra=100$, $Pr=1$, $Re=5$, $\phi=0.6$, and $s_\nu=1$. The grid number $N_y$ varies from 32 to 96. The relative global error of a variable ($\mathbf{u}$ or $T$) is defined by

$$E(\Phi) = \frac{\sqrt{\sum_{\mathbf{x}}|\Phi_A(\mathbf{x})-\Phi_{LB}(\mathbf{x})|^2}}{\sqrt{\sum_{\mathbf{x}}|\Phi_A(\mathbf{x})|^2}} \tag{52}$$

where $\Phi_A$ and $\Phi_{LB}$ represent analytical and numerical solutions, respectively, and the summation is over the entire domain. The relative global errors of the velocity and temperature are plotted logarithmically in Fig. 6, where the symbols denote present results and the lines denote least-square fittings. The results indicate that the present model is approximately second-order accuracy in space.

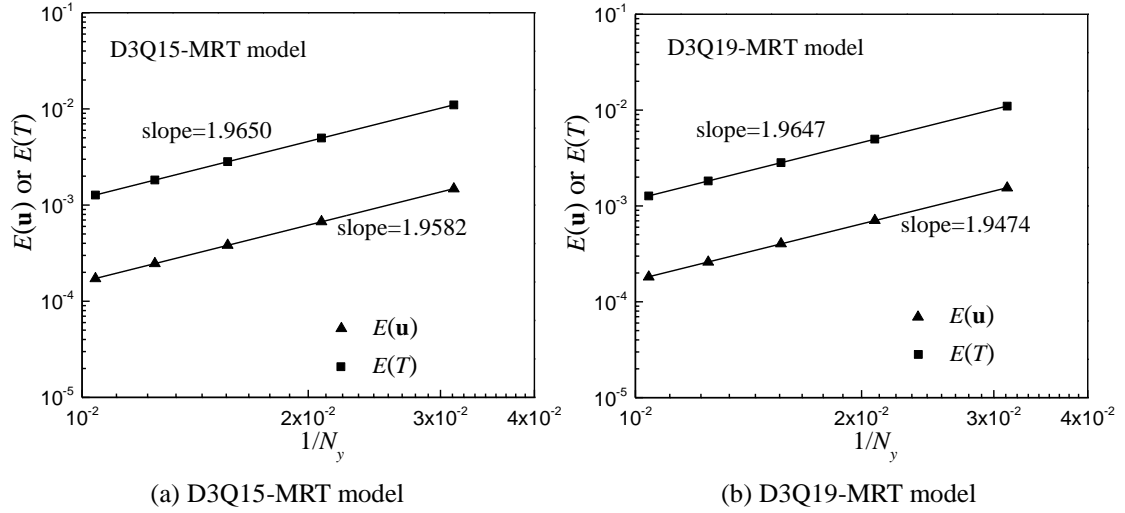

(a) D3Q15-MRT model  (b) D3Q19-MRT model

Fig. 6. Relative global errors of velocity and temperature at different grid numbers

($N_y = 32$, 48, 64, 80, 96).

**4.2 Natural convection in a cubical porous cavity**

In this subsection, the present model is employed to study convection heat transfer in a cubical porous cavity. The schematic of this problem is illustrated in Fig. 7. The length, width, and height of the cubical cavity are $L$, $W$, and $H$ ($L=W=H$), respectively. The left and right walls are kept at constant temperatures $T_h$ and $T_c$ ($T_h > T_c$), respectively, while the other four walls are adiabatic. The buoyancy force is given by $\mathbf{G} = -g\beta(T-T_0)\mathbf{k}$, where $T_0 = (T_h + T_c)/2$ is the reference temperature, and $\mathbf{k}$ is the unit vector in the $z$-direction.

The 3D local Nusselt number $Nu_{\text{local}}(Y,Z)$, the $z$-direction averaged Nusselt number $Nu(Y)$ at the hot wall, and 3D average Nusselt number $Nu_{3D}$ at the hot wall are defined by

$$Nu_{\text{local}}(Y,Z) = -\frac{\partial \theta(Y,Z)}{\partial X}\bigg|_{X=0} \tag{53}$$

$$Nu(Y) = \int_0^1 Nu_{\text{local}}(Y,Z)\,\mathrm{d}Z \tag{54}$$

$$Nu_{3D} = \int_0^1 \int_0^1 Nu_{local}(Y,Z)\,dYdZ = \int_0^1 Nu(Y)\,dY \tag{55}$$

respectively, where $\theta = (T-T_c)/\Delta T$, $\Delta T = T_h - T_c$, and $(X,Y,Z) = (x/L, y/W, z/H)$. The averaged Nusselt number at the hot wall of the symmetry-plane ($Y = 0.5$) $Nu_{mp}$ is given by $Nu_{mp} = Nu(0.5)$.

The relaxation rates $s_v$ and $\zeta_\alpha$ are determined by

$$s_v^{-1} = \frac{1}{2} + \frac{MaJL\sqrt{3Pr}}{c\delta_t\sqrt{Ra}}, \quad \zeta_\alpha^{-1} = \frac{1}{2} + \frac{\Gamma c_s^2(s_v^{-1} - 0.5)}{J\sigma c_{sT}^2 Pr} \tag{56}$$

where $Ma = u_c/c_s$ is the Mach number, in which $u_c = \sqrt{g\beta\Delta TL}$ is the characteristic velocity. For incompressible thermal flows considered in this work, the Mach number $Ma$ should be small and is set to be 0.1.

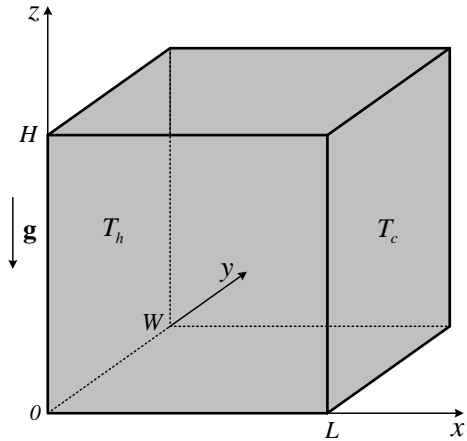

Fig. 7. Natural convection in a cubical porous cavity.

In simulations, the Prandtl number $Pr = 0.71$, the porosity $\phi = 0.4$ and $0.8$, the Rayleigh number $Ra$ ranges from $5\times10^2$ to $10^6$ (Darcy-Rayleigh number $Ra^* = DaRa$ ranges from 50 to $10^3$), and the Darcy number $Da$ ranges from $10^{-3}$ to $10^{-1}$. Considering both the computational time and the accuracy, for $Ra = 10^6$ ($Ra^* = 10^3$), a grid size of $N_x \times N_y \times N_z = 80\times80\times80$ is adopted, for $Ra < 10^6$, a grid size of $N_x \times N_y \times N_z = 64\times64\times64$ is adopted.

In Table 1, the average Nusselt numbers ($Nu_{3D}$) obtained by the present model are compared with the boundary element method (BEM) results [51] of the Brinkman-extended Darcy model ($F_\phi = 0$) for different $Ra^*$ and $Da$ with $\phi = 0.8$. It can be seen that the present results are in good agreement with the BEM results for the whole range of Darcy-Rayleigh and Darcy numbers. In Table 2, the average Nusselt numbers ($Nu_{3D}$) obtained by the present model are compared with the numerical results [51, 52] of the Brinkman-extended Darcy model for different $Da$ with $Ra^* = 10^3$ and $\phi = 0.8$. Very good agreement between these results can be seen from Table 2.

Table 1. Comparison of the present results with the BEM results [51] of the Brinkman-extended Darcy model ($\phi = 0.8$).

| $Ra^*$ | $Da = 10^{-1}$ | | $Da = 10^{-2}$ | | $Da = 10^{-3}$ | |
|---|---|---|---|---|---|---|
| | Ref. [51] | Present | Ref. [51] | Present | Ref. [51] | Present |
| 50 | 1.010 | 1.0095 | 1.216 | 1.2162 | 1.635 | 1.6405 |
| 100 | 1.039 | 1.0372 | 1.533 | 1.5384 | 2.331 | 2.3367 |
| 200 | 1.132 | 1.1292 | 2.029 | 2.0412 | 3.341 | 3.3291 |
| 500 | 1.453 | 1.4534 | 2.920 | 2.9355 | 5.148 | 5.0771 |

Table 2. Comparison of the present results with the numerical results [51, 52] of the Brinkman-extended Darcy model ($Ra^* = 10^3$, $\phi = 0.8$).

| $Da$ | Ref. [51] | Ref. [52] | Present |
|---|---|---|---|
| $10^{-1}$ | 1.855 | 1.854 | 1.8615 |
| $10^{-2}$ | 3.770 | 3.755 | 3.7764 |
| $10^{-3}$ | 6.922 | 6.820 | 6.7085 |

In what follows, based on the generalized non-Darcy model (the inertia coefficient $F_\phi$ is given by Eq. (6)), a parametric study has been conducted for convection heat transfer in the cubical porous cavity for various values of $Ra$, $Da$, and $\phi$. In Fig. 8, the isotherms and contour lines of $u_x$ for $Ra^* = 10^3$ and $\phi = 0.8$ in the symmetry-plane ($Y = 0.5$) are shown. As can be seen from the figure, a decrease in $Da$ leads to an intensification of convection flow inside the cubical porous cavity and as a result to thinner (thermal and velocity) boundary layers near the hot and cold walls. The isotherms and flow patterns in the symmetry-plane are qualitatively similar to those of the 2D convection heat transfer in a square porous cavity. However, the effect of the side walls of 3D convection heat transfer in a cubical porous cavity would be notable, which is to be discussed later.

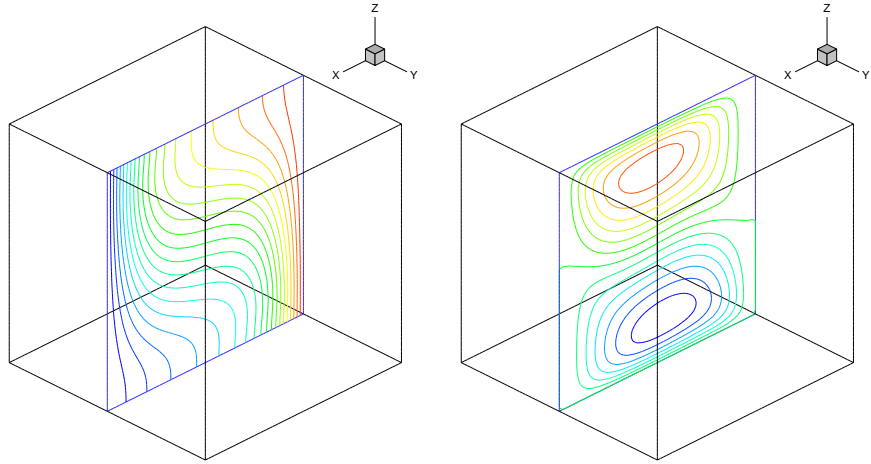

(a)  $Ra = 10^4$, $Da = 10^{-1}$

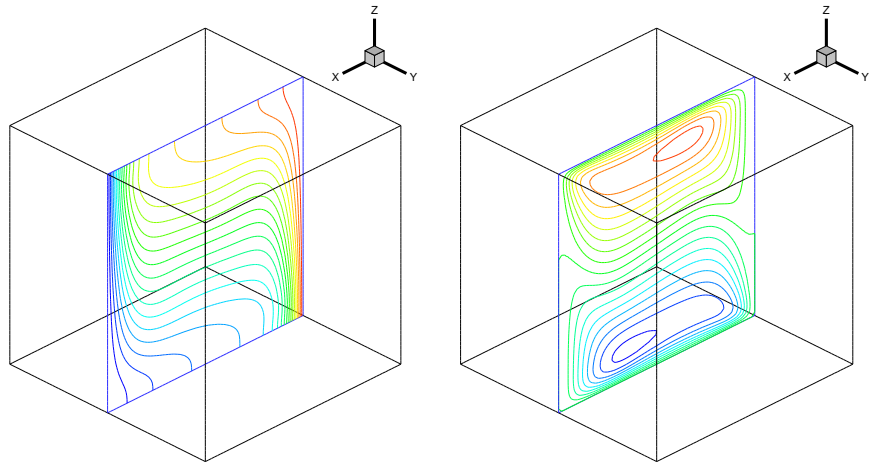

(b)  $Ra = 10^5$, $Da = 10^{-2}$

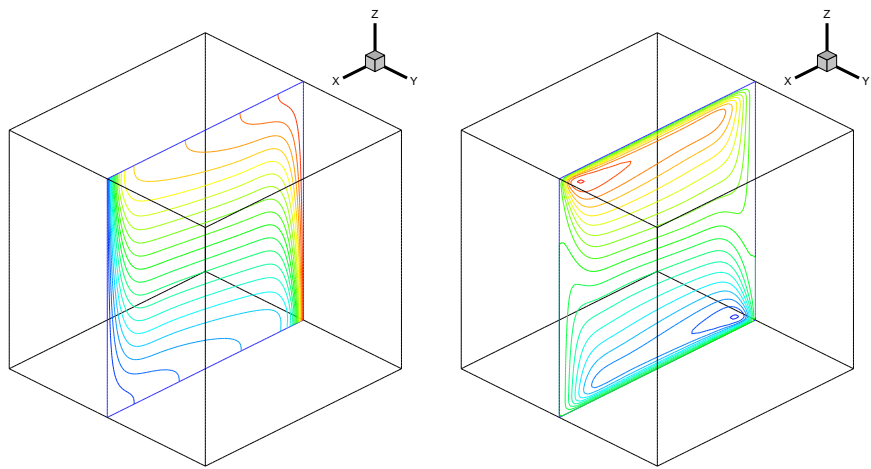

(c)  $Ra = 10^6$, $Da = 10^{-3}$

Fig. 8. Isotherms (left) and contour lines of $u_x$ (right) for $Ra^* = 10^3$ and $\phi = 0.8$ in the symmetry-plane ($Y = 0.5$).

In Table 3, the average Nusselt numbers ($Nu_{3D}$) at the hot wall are presented for various $Ra^*$, $Da$, and $\phi$. The following trends can be observed from the predicted results: (i) for a given $Da$ and $\phi$, $Nu_{3D}$ increases with the increase in $Ra^*$, with the lowest value at $Ra^* = 50$ and highest value at $Ra^* = 10^3$; (ii) for a given $Ra^*$ and $\phi$, $Nu_{3D}$ increases with the decrease in $Da$, and the influence of $Da$ is more pronounced at higher values of $Ra^*$; (iii) for a given $Ra^*$ and $Da$, $Nu_{3D}$ increases with the increase in $\phi$. The influence of $\phi$ on $Nu_{3D}$ is shown in Fig. 9. Clearly, for higher value of $\phi$, the effect of the inertia and nonlinear drag terms are less significant, which leads to higher flow velocities and higher values of $Nu_{3D}$.

Table 3. Average Nusselt numbers of the generalized non-Darcy model.

| $Ra^*$ | $\phi = 0.4$ | | | $\phi = 0.8$ | | |
|---|---|---|---|---|---|---|
| | $Da = 10^{-1}$ | $Da = 10^{-2}$ | $Da = 10^{-3}$ | $Da = 10^{-1}$ | $Da = 10^{-2}$ | $Da = 10^{-3}$ |
| 50 | 1.0029 | 1.0990 | 1.4509 | 1.0094 | 1.2000 | 1.5968 |
| 100 | 1.0109 | 1.2730 | 1.9409 | 1.0361 | 1.4925 | 2.2337 |
| 200 | 1.0403 | 1.5880 | 2.6316 | 1.1233 | 1.9535 | 3.1356 |
| 500 | 1.1836 | 2.2011 | 3.8173 | 1.4291 | 2.7899 | 4.6877 |
| 1000 | 1.4289 | 2.7916 | 4.9193 | 1.8124 | 3.5836 | 6.1256 |

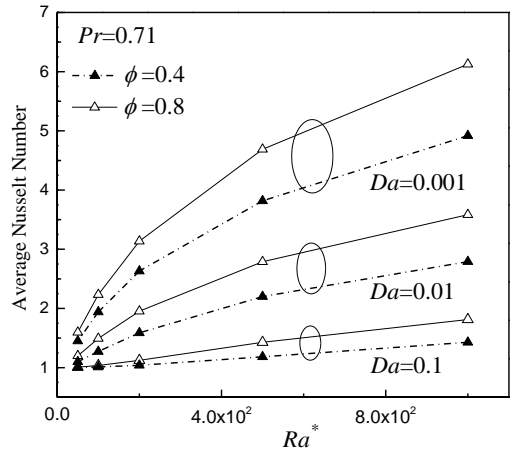

Fig. 9. The influence of $\phi$ on $Nu_{3D}$ of the generalized non-Darcy model.

The influence of the nonlinear drag force on $Nu_{3D}$ is shown in Fig. 10. In the figure, BD model represents the Brinkman-extended Darcy model, and BFD model represents the generalized non-Darcy model. Form the figure we can observe that for $Da=0.1$, the results are almost the same for the BD and BFD models, the influence of the nonlinear drag force can be neglected. While for $Da<0.1$ and $Ra^*>100$, the nonlinear drag force becomes significant and reduces the overall heat transfer, which leads to smaller values of $Nu_{3D}$. For non-Darcy flows ($10^{-4}<Da<10^{-1}$) with relatively large $Ra^*$, it is essential to consider the influence of the nonlinear drag force.

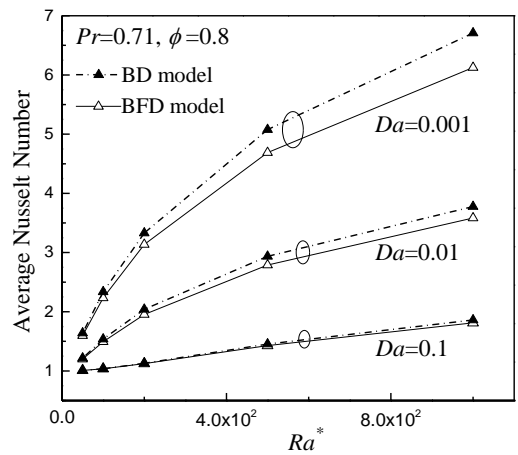

Fig.10. The influence of the nonlinear drag force on $Nu_{3D}$.

Finally, the effect of the side walls of 3D convection heat transfer in the cubical porous cavity is studied. Here we only consider the case with $Ra=10^5$, $Da=10^{-2}$, and $\phi=0.8$ without loss of generality. In Fig. 11, the average Nusselt number along the *y*-direction at the hot wall ($Nu(Y)$) is shown. From the figure we can observe that $Nu(Y)$ increases as $Y\to 0.5$, and the maximum value occurs at $Y=0.5$, i.e., $Nu_{max}(Y)=Nu_{mp}=3.7377$. The 3D average Nusselt number $Nu_{3D}$ at the hot wall is 3.5836 (see Table 2), while the 2D average Nusselt number $Nu_{2D}$ at the hot wall of the 2D square porous cavity is 3.7054 (obtained by the 2D DDF-MRT model [30] based on a grid size of $N_x \times N_y = 100\times 100$). Owing to the effect of the side walls, the 2D result overestimates the 3D effective heat transfer, but underestimates the symmetry-plane effective heat transfer. With respect to $Nu_{mp}$, the 3D heat transfer is always weaker.

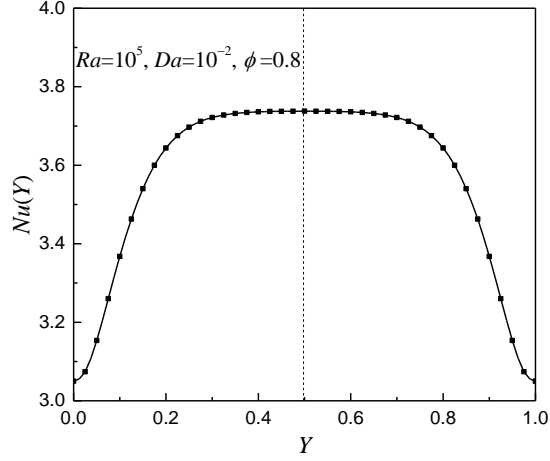

Fig.11. The distribution of the *z*-direction averaged Nusselt number ( $X = 0$ ).

**4.3 Two-region conduction melting in a semi-infinite space**

In this subsection, numerical simulations of two-region conduction melting in a semi-infinite space (see Fig. 12) are carried out to validate the enthalpy-based D3Q7-MRT model. In this problem, the initial temperature ( $T_i$ ) of the phase change material (PCM) is below the melting point of the PCM ( $T_m$ ), the temperatures of both the liquid and solid phases are unknown and must be determined. This is the so-called *two-region* conduction melting problem. At time $t = 0$, a constant temperature $T_h$ ( $T_h > T_m$ ) is imposed on the left wall ( $x = 0$ ) and maintained at that temperature for $t > 0$. As a results, the melting starts at $x = 0$ and the interface moves in the positive *x*-direction. The temperatures of the liquid and solid phases are given by [56]

$$T(x,t) = T_h - \frac{(T_h - T_m)\,\mathrm{erf}\left[x/\left(2\sqrt{\alpha_l t}\right)\right]}{\mathrm{erf}(\eta)}, \quad 0 < x \le x_m(t) \quad \text{(liquid region)} \tag{57}$$

$$T(x,t) = T_i + \frac{(T_m - T_i)\,\mathrm{erfc}\left[x/\left(2\sqrt{\alpha_s t}\right)\right]}{\mathrm{erfc}\left(\eta\sqrt{\alpha_l/\alpha_s}\right)}, \quad x > x_m(t) \quad \text{(solid region)} \tag{58}$$

respectively, where $x_m(t) = 2\eta\sqrt{\alpha_l t}$ is the location of the solid-liquid phase interface, $\mathrm{erf}(\eta) = \frac{2}{\sqrt{\pi}}\int_0^\eta e^{-\vartheta^2}\,\mathrm{d}\vartheta$ is the error function, and $\mathrm{erfc}(x) = 1 - \mathrm{erf}(x)$ is the complementary error function. The parameter $\eta$ can be determined by the following transcendental equation [56]

$$\frac{e^{-\eta^2}}{\text{erf}(\eta)} + \frac{k_s}{k_l}\left(\frac{\alpha_l}{\alpha_s}\right)^{1/2} \frac{T_i - T_m}{T_h - T_m} \frac{e^{-\eta^2(\alpha_l/\alpha_s)}}{\text{erfc}(\eta\sqrt{\alpha_l/\alpha_s})} = \frac{\eta L_a \sqrt{\pi}}{c_{pl}(T_h - T_m)} \qquad (59)$$

In simulations, the parameters are set as follows: $T_h = 1$, $T_m = 0$, $T_i = -1$, $\alpha_s = 0.02$, $\phi = 1$, $\sigma = 1$, $c_{pl} = c_{ps} = 1$, $St = c_{pl}\Delta T/L_a = 1$ ($\Delta T = T_h - T_m$). A grid size of $N_x \times N_y \times N_z = 400 \times 6 \times 6$ is employed, and the thermal diffusivity ratio $\alpha_l/\alpha_s$ varies from 1 to 12. To simulate such one-dimensional conduction melting problem, the periodic boundary conditions are imposed in the y- and z-directions, and the velocity field is set to be zero consistently ($\mathbf{u} = 0$). The relaxation rate $\zeta_\alpha$ is determined by $\zeta_\alpha^{-1} = 0.5 + \alpha_e/(c_{sT}^2 \delta_t)$, where $\alpha_e = \alpha_l f_l + \alpha_s(1 - f_l)$. In Fig. 13, the temperature profiles for different values of thermal diffusivity ratio at $Fo = 0.01$ are plotted ($Fo = t\alpha_s/L^2$). It can be observed that the present results are in good agreement with the analytical results. The locations of the solid-liquid phase interface for different values of thermal diffusivity ratio and $Fo$ are plotted in Fig. 14. Good agreement can be observed again between the present results and analytical results.

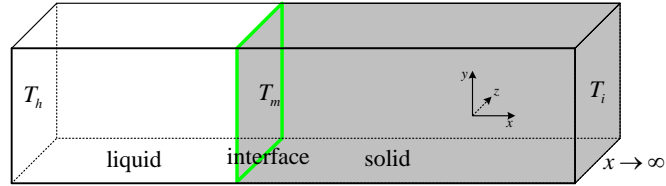

Fig. 12. Schematic of the two-region conduction melting in a semi-infinite space.

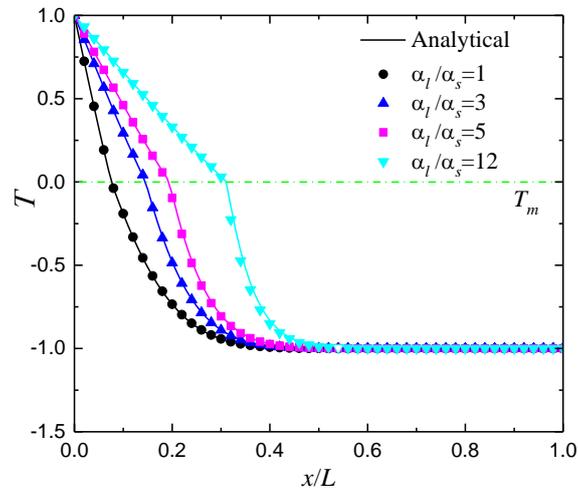

Fig. 13. Temperature profiles for different values of thermal diffusivity ratio at $Fo$=0.01.

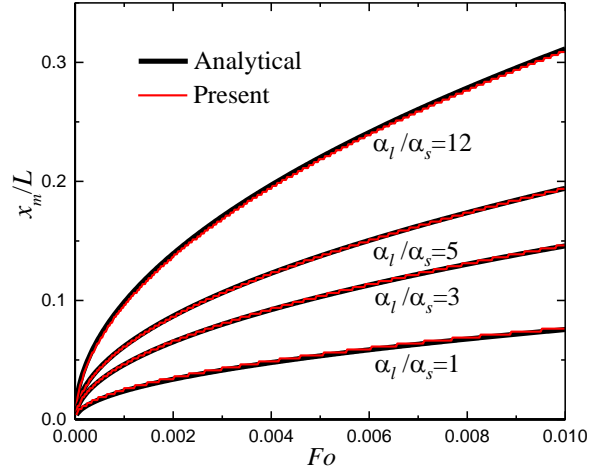

Fig. 14. Locations of the solid-liquid phase interface for different values of thermal diffusivity ratio and *Fo*.

## 5. Conclusions

In this paper, a 3D DDF-MRT model is presented for convection heat transfer in porous media at the REV scale. The DDF-MRT model consists of two different MRT-LB models: an MRT-LB model of the density distribution function with the D3Q19 lattice (or D3Q15 lattice) is proposed to simulate the flow field based on the generalized non-Darcy model, and an MRT-LB model of the temperature distribution function with the D3Q7 lattice is proposed to simulate the temperature filed. The key point of the model is to include the porosity into the equilibrium moments and add a forcing term to the MRT-LB equation of the flow field to account for the linear (Darcy's term) and nonlinear (Forchheimer's term) drag forces of the porous matrix based on the generalized non-Darcy model.

The present model is first employed to simulate mixed convection flow in a porous channel. It is found that the present results agree well with the analytical solutions, and the results demonstrate that the present model is approximately second-order accuracy in space. Then the present model is employed to simulate convection heat transfer in a cubic porous cavity in the non-Darcy flow regime. The results demonstrate that the present model has the applicability to solve 3D convection heat transfer problems in porous media. In addition, an enthalpy-based DDF-MRT model for 3D solid-liquid phase change with convection heat transfer in porous media is also presented.

## Appendix A. Transformation matrices

For the D3Q15-MRT model, the transformation matrix **M** is given by ($c=1$) [49]

$$\mathbf{M} = \begin{bmatrix}
1 & 1 & 1 & 1 & 1 & 1 & 1 & 1 & 1 & 1 & 1 & 1 & 1 & 1 & 1 \\
-2 & -1 & -1 & -1 & -1 & -1 & -1 & 1 & 1 & 1 & 1 & 1 & 1 & 1 & 1 \\
16 & -4 & -4 & -4 & -4 & -4 & -4 & 1 & 1 & 1 & 1 & 1 & 1 & 1 & 1 \\
0 & 1 & -1 & 0 & 0 & 0 & 0 & 1 & -1 & 1 & -1 & 1 & -1 & 1 & -1 \\
0 & -4 & 4 & 0 & 0 & 0 & 0 & 1 & -1 & 1 & -1 & 1 & -1 & 1 & -1 \\
0 & 0 & 0 & 1 & -1 & 0 & 0 & 1 & 1 & -1 & -1 & 1 & 1 & -1 & -1 \\
0 & 0 & 0 & -4 & 4 & 0 & 0 & 1 & 1 & -1 & -1 & 1 & 1 & -1 & -1 \\
0 & 0 & 0 & 0 & 0 & 1 & -1 & 1 & 1 & 1 & 1 & -1 & -1 & -1 & -1 \\
0 & 0 & 0 & 0 & 0 & -4 & 4 & 1 & 1 & 1 & 1 & -1 & -1 & -1 & -1 \\
0 & 2 & 2 & -1 & -1 & -1 & -1 & 0 & 0 & 0 & 0 & 0 & 0 & 0 & 0 \\
0 & 0 & 0 & 1 & 1 & -1 & -1 & 0 & 0 & 0 & 0 & 0 & 0 & 0 & 0 \\
0 & 0 & 0 & 0 & 0 & 0 & 0 & 1 & -1 & -1 & 1 & 1 & -1 & -1 & 1 \\
0 & 0 & 0 & 0 & 0 & 0 & 0 & 1 & 1 & -1 & -1 & -1 & -1 & 1 & 1 \\
0 & 0 & 0 & 0 & 0 & 0 & 0 & 1 & -1 & 1 & -1 & -1 & 1 & -1 & 1 \\
0 & 0 & 0 & 0 & 0 & 0 & 0 & 1 & -1 & -1 & 1 & -1 & 1 & 1 & -1
\end{bmatrix} \quad (A1)$$

For the D3Q19-MRT model, the transformation matrix **M** is given by ($c=1$) [49]

$$\mathbf{M} = \begin{bmatrix}
1 & 1 & 1 & 1 & 1 & 1 & 1 & 1 & 1 & 1 & 1 & 1 & 1 & 1 & 1 & 1 & 1 & 1 & 1 \\
-30 & -11 & -11 & -11 & -11 & -11 & -11 & 8 & 8 & 8 & 8 & 8 & 8 & 8 & 8 & 8 & 8 & 8 & 8 \\
12 & -4 & -4 & -4 & -4 & -4 & -4 & 1 & 1 & 1 & 1 & 1 & 1 & 1 & 1 & 1 & 1 & 1 & 1 \\
0 & 1 & -1 & 0 & 0 & 0 & 0 & 1 & -1 & 1 & -1 & 1 & -1 & 1 & -1 & 0 & 0 & 0 & 0 \\
0 & -4 & 4 & 0 & 0 & 0 & 0 & 1 & -1 & 1 & -1 & 1 & -1 & 1 & -1 & 0 & 0 & 0 & 0 \\
0 & 0 & 0 & 1 & -1 & 0 & 0 & 1 & 1 & -1 & -1 & 0 & 0 & 0 & 0 & 1 & -1 & 1 & -1 \\
0 & 0 & 0 & -4 & 4 & 0 & 0 & 1 & 1 & -1 & -1 & 0 & 0 & 0 & 0 & 1 & -1 & 1 & -1 \\
0 & 0 & 0 & 0 & 0 & 1 & -1 & 0 & 0 & 0 & 0 & 1 & 1 & -1 & -1 & 1 & 1 & -1 & -1 \\
0 & 0 & 0 & 0 & 0 & -4 & 4 & 0 & 0 & 0 & 0 & 1 & 1 & -1 & -1 & 1 & 1 & -1 & -1 \\
0 & 2 & 2 & -1 & -1 & -1 & -1 & 1 & 1 & 1 & 1 & 1 & 1 & 1 & 1 & -2 & -2 & -2 & -2 \\
0 & -4 & -4 & 2 & 2 & 2 & 2 & 1 & 1 & 1 & 1 & 1 & 1 & 1 & 1 & -2 & -2 & -2 & -2 \\
0 & 0 & 0 & 1 & 1 & -1 & -1 & 1 & 1 & 1 & 1 & -1 & -1 & -1 & -1 & 0 & 0 & 0 & 0 \\
0 & 0 & 0 & -2 & -2 & 2 & 2 & 1 & 1 & 1 & 1 & -1 & -1 & -1 & -1 & 0 & 0 & 0 & 0 \\
0 & 0 & 0 & 0 & 0 & 0 & 0 & 1 & -1 & -1 & 1 & 0 & 0 & 0 & 0 & 0 & 0 & 0 & 0 \\
0 & 0 & 0 & 0 & 0 & 0 & 0 & 0 & 0 & 0 & 0 & 0 & 0 & 0 & 0 & 1 & -1 & -1 & 1 \\
0 & 0 & 0 & 0 & 0 & 0 & 0 & 0 & 0 & 0 & 0 & 1 & -1 & -1 & 1 & 0 & 0 & 0 & 0 \\
0 & 0 & 0 & 0 & 0 & 0 & 0 & 1 & -1 & 1 & -1 & -1 & 1 & -1 & 1 & 0 & 0 & 0 & 0 \\
0 & 0 & 0 & 0 & 0 & 0 & 0 & -1 & -1 & 1 & 1 & 0 & 0 & 0 & 0 & 1 & -1 & 1 & -1 \\
0 & 0 & 0 & 0 & 0 & 0 & 0 & 0 & 0 & 0 & 0 & 1 & 1 & -1 & -1 & -1 & -1 & 1 & 1
\end{bmatrix} \quad (A2)$$

**Appendix B. Enthalpy-based D3Q7-MRT model**

For solid-liquid phase change with convection heat transfer in porous media, under the LTE condition, the enthalpy-based energy governing equation can be written as [53]

$$\frac{\partial}{\partial t}\left\{\phi\left[f_l\rho_l En_l+(1-f_l)\rho_s En_s\right]+(1-\phi)\rho_m En_m\right\}+\nabla\cdot(\rho_l En_l \mathbf{u})=\nabla\cdot(k_e\nabla T) \tag{B1}$$

where $f_l$ is the fraction of the liquid PCM in the pore space, $En_l$, $En_s$, and $En_m$ are enthalpy of the liquid PCM, solid PCM, and porous matrix, respectively, $k_e$ is the effective thermal conductivity, and $L_a$ is the latent heat of phase change. The subscripts $l$, $s$, and $m$ refer to the properties of the liquid PCM, solid PCM, and porous matrix, respectively. $En_l$, $En_s$, and $En_m$ are defined by

$$En_l=c_{pl}T+f_l L_a, \quad En_s=c_{ps}T, \quad En_m=c_{pm}T \tag{B2}$$

Substituting Eq. (B2) into Eq. (B1), after a few steps, the following energy governing equation can be derived

$$\frac{\partial}{\partial t}\left(\sigma c_{pl}T+\phi f_l L_a\right)+\nabla\cdot\left(c_{pl}T\mathbf{u}\right)=\nabla\cdot\left(\frac{k_e}{\rho_l}\nabla T\right)-\nabla\cdot(f_l L_a \mathbf{u}) \tag{B3}$$

where $\rho_l=\rho_s=\rho_f\approx\rho_0$, $\sigma$ is the heat capacity ratio (ratio between mean heat capacity of the mixture and liquid heat capacity) [53]

$$\sigma=\frac{\phi\left[f_l\rho_l c_{pl}+(1-f_l)\rho_s c_{ps}\right]+(1-\phi)\rho_m c_{pm}}{\rho_l c_{pl}} \tag{B4}$$

In liquid region, $\sigma_l=[\phi\rho_l c_{pl}+(1-\phi)\rho_m c_{pm}]/(\rho_l c_{pl})$; in solid region, $\sigma_s=[\phi\rho_s c_{ps}+(1-\phi)\rho_m c_{pm}]/(\rho_l c_{pl})$.

Note that the last term in Eq. (B3), i.e., $\nabla\cdot(L_a f_l \mathbf{u})$, which is induced by the flow in the mushy zone, can be neglected for isothermal solid-liquid phase change [53-55]. Then by introducing an effective enthalpy $En_e=\sigma c_{pl}T+\phi f_l L_a$ [35], the following effective-enthalpy-based energy governing equation can be obtained

$$\frac{\partial En_e}{\partial t}+\nabla\cdot\left(c_{pl}T\mathbf{u}\right)=\nabla\cdot\left(\frac{k_e}{\rho_l}\nabla T\right) \tag{B5}$$

For the enthalpy-based D3Q7-MRT model, the equilibrium moments $\{n_i^{eq}\}$ are defined as

$$\mathbf{n}^{eq} = \left(En_e, c_{pl}Tu_x, c_{pl}Tu_y, c_{pl}Tu_z, \varpi c_{p,\text{ref}}T, 0, 0\right)^T \tag{B6}$$

where $\varpi \in (0,1)$, $c_{p,\text{ref}}$ is the reference specific heat.

The effective enthalpy $En_e$ is computed by

$$En_e = n_0 = \sum_i g_i \tag{B7}$$

The relationship between effective enthalpy $En_e$ and temperature $T$ is given by

$$T = \begin{cases} En_e/(\sigma_s c_{pl}), & En_e \leq En_{e,s} \\ T_s + \dfrac{En_e - En_{e,s}}{En_{e,l} - En_{e,s}}(T_l - T_s), & En_{e,s} < En_e < En_{e,l} \\ T_l + (En_e - En_{e,l})/(\sigma_l c_{pl}), & En_e \geq En_{e,l} \end{cases} \tag{B8}$$

where $T_s$ and $T_l$ ($T_s \leq T_l$) are the solidus and liquidus temperatures, respectively, $En_{e,s}$ ($En_{e,s} = \sigma_s c_{pl} T_s$) and $En_{e,l}$ ($En_e = \sigma_l c_{pl} T_l + \phi f_l L_a$) are the effective enthalpy values corresponding to $T_s$ and $T_l$, respectively. The liquid fraction $f_l$ can be calculated by

$$f_l = \begin{cases} 0, & En_e \leq En_{e,s} \\ \dfrac{En_e - En_{e,s}}{En_{e,l} - En_{e,s}}, & En_{e,s} < En_e < En_{e,l} \\ 1, & En_e \geq En_{e,l} \end{cases} \tag{B9}$$

Through the Chapman-Enskog analysis, the following macroscopic equation can be obtained

$$\frac{\partial En_e}{\partial t} + \nabla \cdot (c_{pl}T\mathbf{u}) = \nabla \cdot \left\{\delta_t\left(\zeta_\alpha^{-1} - 0.5\right)\left[\nabla\left(c_{sT}^2 c_{p,\text{ref}}T\right) + \epsilon\partial_{t_1}(c_{pl}T\mathbf{u})\right]\right\} \tag{B10}$$

For incompressible thermal flows, the deviation term $\nabla \cdot \left[\delta_t\left(\zeta_\alpha^{-1} - 0.5\right)\epsilon\partial_{t_1}(c_{pl}T\mathbf{u})\right]$ can be neglected, then the effective-enthalpy-based energy governing equation (B5) can be recovered with

$$\alpha_e = c_{sT}^2\left(\frac{1}{\zeta_\alpha} - \frac{1}{2}\right)\delta_t = \frac{k_e}{\rho_l c_{p,\text{ref}}} \tag{B11}$$

The equilibrium distribution function $g_i^{eq}$ ($\mathbf{g}^{eq} = \mathbf{N}^{-1}\mathbf{n}^{eq}$) in the velocity space is given by

$$g_i^{eq} = \begin{cases} En_e - \varpi c_{p,\text{ref}}T, & i = 0 \\ \dfrac{1}{6}\varpi c_{p,\text{ref}}T + \dfrac{1}{2}(\mathbf{e}_i \cdot \mathbf{u})c_{pl}T, & i = 1 \sim 6 \end{cases} \tag{B12}$$

As did in Ref. [55], the reference specific heat $c_{p,\text{ref}}$ is introduced into the model. Note that the choice of $c_{p,\text{ref}}$ can be arbitrary. In the enthalpy-based D3Q7-MRT model, $c_{p,\text{ref}}$ can be chosen as the harmonic mean of the specific heats, i.e.,

$$c_{p,\text{ref}} = \frac{2}{\dfrac{1}{c_{pl}} + \dfrac{1}{c_{ps}}} \quad \text{or} \quad c_{p,\text{ref}} = \frac{3}{\dfrac{1}{c_{pl}} + \dfrac{1}{c_{ps}} + \dfrac{1}{c_{pm}}} \tag{B13}$$

The reference specific heat $c_{p,\text{ref}}$ given by Eq. (B13) keeps unvaried in space ($c_{pl}$, $c_{ps}$, and $c_{pm}$ are constants). For solid-liquid phase change with convection heat transfer in porous media, the collision process of the MRT-LB model of the flow field needs to be considered for the liquid phase only ($f_l \geq 0.5$).

## References

[1] J. Bear, Dynamics of Fluids in Porous Media, Elsevier, New York, 1972.

[2] P. Cheng, Heat transfer in geothermal systems, Adv. Heat Transfer 14 (1978) 1-105.

[3] D.B. Ingham, I. Pop (Eds.), Transport Phenomena in Porous Media, Pergamon Press, Danvers, 1998.

[4] D.A. Nield, A. Bejan, Convection in Porous Media, 4th ed., Springer, New York, 2013.

[5] G.R. McNamara, G. Zanetti, Use of the Boltzmann equation to simulate lattice-gas automata, Phys. Rev. Lett. 61(20) (1988) 2332-2335.

[6] F.J. Higuera, S. Succi, R. Benzi, Lattice gas dynamics with enhanced collisions, Europhys. Lett. 9(7) (1989) 345-349.

[7] Y.H. Qian, D. d'Humières, P. Lallemand, Lattice BGK models for Navier-Stokes equation, Europhys. Lett. 17(6) (1992) 479-484.

[8] S. Chen, G.D. Doolen, Lattice Boltzmann method for fluid flows, Annu. Rev. Fluid Mech. 30(1) (1998) 329-364.

[9] S. Succi, The Lattice Boltzmann Equation for Fluid Dynamics and Beyond, Clarendon Press, Oxford, 2001.

[10] Y.L. He, Y. Wang, Q. Li, Lattice Boltzmann Method: Theory and Applications, Science Press, Beijing, 2009.

[11] S. Gong, P. Cheng, Lattice Boltzmann simulation of periodic bubble nucleation, growth and departure from a heated surface in pool boiling, Int. J. Heat Mass Transfer 64 (2013) 122-132.

[12] Q. Li, Q.J. Kang, M.M. Francois, Y.L. He, K.H. Luo, Lattice Boltzmann modeling of boiling heat transfer: The boiling curve and the effects of wettability, Int. J. Heat Mass Transfer 85 (2015) 787-796.

[13] S. Succi, Lattice Boltzmann 2038, Europhys. Lett. 109(5) (2015) 50001.

[14] Q. Li, K.H. Luo, Q.J. Kang, Y.L. He, Q. Chen, Q. Liu, Lattice Boltzmann methods for multiphase flow and phase-change heat transfer, Prog. Energy. Combust. Sci. 52 (2016) 62-105.

[15] U. Frisch, B. Hasslacher, Y. Pomeau, Lattice-gas automata for the Navier-Stokes equation, Phys. Rev. Lett. 56 (1986) 1505-1508.

[16] S. Succi, E. Foti, F. Higuera, Three-dimensional flows in complex geometries with the lattice Boltzmann method, Europhys. Lett. 10(5) (1989) 433.

[17] D. Zhang, R. Zhang, S. Chen, W.E. Soll, Pore scale study of flow in porous media: Scale dependency, REV, and statistical REV, Geophys. Res. Let. 27(8) (2000) 1195-1198.

[18] G.H. Tang, W.Q. Tao, Y. L. He, Gas slippage effect on microscale porous flow using the lattice Boltzmann method, Phys. Rev. E 72(5) (2005) 056301.

[19] Q. Kang, P.C. Lichtner, D. Zhang, An improved lattice Boltzmann model for multicomponent reactive transport in porous media at the pore scale, Water Resour. Res. 43(12) (2007).

[20] L. Hao, P. Cheng, Lattice Boltzmann simulations of anisotropic permeabilities in carbon paper gas diffusion layers, J. Power Sources 186(1) (2009) 104-114.

[21] L. Chen, H.B Luan, Y.L He, W.Q. Tao, Pore-scale flow and mass transport in gas diffusion layer of proton exchange membrane fuel cell with interdigitated flow fields, Int. J. Therm. Sci. 51 (2012) 132-144.

[22] M.A.A. Spaid, F.R. Phelan Jr, Lattice Boltzmann methods for modeling microscale flow in fibrous porous media, Phys. Fluids 9(9) (1997) 2468-2474.

[23] O. Dardis, J. McCloskey, Lattice Boltzmann scheme with real numbered solid density for the simulation of flow in porous media, Phys. Rev. E 57 (1998) 4834-4837.

[24] Q. Kang, D. Zhang, S. Chen, Unified lattice Boltzmann method for flow in multiscale porous media, Phys. Rev. E 66 (2002) 056307.

[25] Z. Guo, T.S. Zhao, Lattice Boltzmann model for incompressible flows through porous media, Phys. Rev. E 66 (2002) 036304.

[26] L. Chen, W. Fang, Q. Kang, J.D.H. Hyman, H.S. Viswanathan, W.Q. Tao, Generalized lattice Boltzmann model for flow through tight porous media with Klinkenberg's effect, Phys. Rev. E 91(3) (2015) 033004.

[27] Z. Guo, T.S. Zhao, A lattice Boltzmann model for convection heat transfer in porous media, Numer. Heat Transfer B 47 (2005) 157-177.

[28] T. Seta, E. Takegoshi, K. Okui, Lattice Boltzmann simulation of natural convection in porous media, Math. Comput. Simulat. 72(2) (2006) 195-200.

[29] F. Rong, Z. Guo, Z. Chai, B. Shi, A lattice Boltzmann model for axisymmetric thermal flows through porous media, Int. J. Heat Mass Transfer 53(23) (2010) 5519-5527.

[30] Q. Liu, Y.L. He, Q. Li, W.Q. Tao, A multiple-relaxation-time lattice Boltzmann model for convection heat transfer in porous media, Int. J. Heat Mass Transfer 73 (2014) 761-775.

[31] L. Wang, J. Mi, Z. Guo, A modified lattice Bhatnagar-Gross-Krook model for convection heat transfer in porous media, Int. J. Heat Mass Transfer 94 (2016) 269-291.

[32] D. Gao, Z. Chen, Lattice Boltzmann simulation of natural convection dominated melting in a rectangular cavity filled with porous media, Int. J. Therm. Sci. 50 (2011) 493-501.

[33] D. Gao, Z. Chen, L. Chen, A thermal lattice Boltzmann model for natural convection in porous media under local thermal non-equilibrium conditions, Int. J. Heat Mass Transfer 70 (2014) 979-989.

[34] Q. Liu, Y.L. He, Double multiple-relaxation-time lattice Boltzmann model for solid-liquid phase change with natural convection in porous media, Physica A 438 (2015) 94-106.

[35] W. Wu, S. Zhang, S. Wang, A novel lattice Boltzmann model for the solid-liquid phase change with the convection heat transfer in the porous media, Int. J. Heat Mass Transfer 104 (2017) 675-687.

[36] P. Nithiarasu, K.N. Seetharamu, T. Sundararajan, Natural convective heat transfer in a fluid saturated variable porosity medium, Int. J. Heat Mass Transfer 40(16) (1997) 3955-3967.

[37] D. d'Humières, Generalized lattice-Boltzmann equations, in: B.D. Shizgal, D.P. Weaver (Eds.), Rarefied Gas Dynamics: Theory and Simulations, in: Prog. Astronaut. Aeronaut., Vol. 159, AIAA, Washington, DC, 1992, pp. 450-458.

[38] P. Lallemand, L.-S. Luo, Theory of the lattice Boltzmann method: Dispersion, dissipation, isotropy, Galilean invariance, and stability, Phys. Rev. E 61 (2000) 6546-6562.

[39] C.T. Hsu, P. Cheng, Thermal dispersion in a porous medium, Int. J. Heat Mass Transfer 33(8)

(1990) 1587-1597.

[40] S. Ergun, Fluid flow through packed columns, Chem. Eng. Prog. 48 (1952) 89-94.

[41] K. Vafai, Convective flow and heat transfer in variable-porosity media, J. Fluid Mech. 147 (1984) 233-259.

[42] A. Mezrhab, M.A. Moussaoui, M. Jami, H. Naji, M. Bouzidi, Double MRT thermal lattice Boltzmann method for simulating convective flows, Phys. Lett. A 374(34) (2010) 3499-3507.

[43] J. Wang, D. Wang, P. Lallemand, L.-S. Luo, Lattice Boltzmann simulations of thermal convective flows in two dimensions, Comput. Math. Appl. 66(2) (2013) 262-286.

[44] Q. Liu, Y.L. He, D. Li, Q. Li, Non-orthogonal multiple-relaxation-time lattice Boltzmann method for incompressible thermal flows, Int. J. Heat Mass Transfer 102 (2016) 1334-1344.

[45] H. Wu, J. Wang, Z. Tao, Passive heat transfer in a turbulent channel flow simulation using large eddy simulation based on the lattice Boltzmann method framework, Int. J. Heat Fluid Flow 32 (2011) 1111-1119.

[46] H.H. Xia, G.H. Tang, Y. Shi, W.Q. Tao, Simulation of heat transfer enhancement by longitudinal vortex generators in dimple heat exchangers, Energy 74 (2014) 27-36.

[47] K.N. Premnath, J. Abraham, Three-dimensional multi-relaxation time (MRT) lattice-Boltzmann models for multiphase flow, J. Comput. Phys. 224(2) (2007) 539-559.

[48] Q. Li, Y.L. He, G.H. Tang, W.Q. Tao, Improved axisymmetric lattice Boltzmann scheme, Phys. Rev. E 81 (2010) 056707.

[49] D. d'Humières, I. Ginzburg, M. Krafczyk, P. Lallemand, L.S. Luo, Multiple-relaxation-time lattice Boltzmann models in three dimensions, Phil. Trans. R. Soc. Lond. A 360 (1792) (2002) 437-451.

[50] Z.L. Guo, C.G. Zheng, B.C. Shi, Non-equilibrium extrapolation method for velocity and pressure

boundary conditions in the lattice Boltzmann method, Chin. Phys. 11(4) (2002) 366-374.

[51] J. Kramer, J. Ravnik, R. Jecl, L. Škerget, Simulation of 3D flow in porous media by boundary element method, Eng. Anal. Bound. Elem. 35(12) (2011) 1256-1264.

[52] M. Sheremet, T. Grosan, I. Pop, Natural convection in a cubical porous cavity saturated with nanofluid using Tiwari and Das' nanofluid model, J. Porous Media 18(6) (2015) 585-596.

[53] C. Beckermann, R. Viskanta, Natural convection solid/liquid phase change in porous media, Int. J. Heat Mass Transfer 31 (1988) 35-46.

[54] S. Chakraborty, D. Chatterjee, An enthalpy-based hybrid lattice-Boltzmann method for modelling solid-liquid phase transition in the presence of convective transport, J. Fluid Mech. 592 (2007) 155-176.

[55] R. Huang, H. Wu, Phase interface effects in the total enthalpy-based lattice Boltzmann model for solid-liquid phase change, J. Computat. Phys. 294 (2015) 346-362.

[56] H. Hu, S.A. Argyropoulos, Mathematical modeling of solidification and melting: a review, Model. Simul. Mater. Sci. Eng. 4 (1996) 371-396.